
\tolerance=10000
\magnification=\magstep1

\def\tallbox#1#2{{\setbox0\hbox{#2}\dimen0=\ht0\dimen1=\dp0%
\advance\dimen0 by #1truemm\advance\dimen1 by #1truemm%
\hbox{\vrule height\dimen0 depth\dimen1 width0pt}\box0}}


\def\vec#1{{\bf#1}}
\def\<{\langle}
\def\>{\rangle}

\def\Borel{M_B^2}
\def\Borels{M_B^4}
\def\PI{{\varphi}}
\def\U{q_1}
\def\D{q_2}
\def\u{1}
\def\d{2}
\def\PA{Present address: Suzuka University of Medical Science and Technology,
1001-1 Kishioka, Suzuka, Mie 510-02, Japan}

\hsize=15.7 true cm
\hfill \vbox{
\hbox{} }
\baselineskip=15pt plus 0.5pt minus 0.1pt
\rightline{INS-Rep.-1059}
\rightline{RCNP Preprint 073}
\vskip 20pt
\centerline{\bf  Pion-Nucleon and Kaon-Nucleon Scattering Lengths in QCD Sum
Rules}
\vskip 20pt
\centerline{\sl   Y.~Kondo\footnote{$^1$}{\PA}, O.~Morimatsu}
\vskip 8pt
\centerline{\sl Institute for Nuclear Study, University of Tokyo,}
\centerline{\sl Midori-cho, Tanashi, Tokyo 188}
\vskip 8pt
\centerline{\sl   and}
\vskip 8pt
\centerline{\sl   Y.~Nishino}
\vskip 8pt
\centerline{\sl Research Center for Nuclear Physics, Osaka University,}
\centerline{\sl Mihogaoka, Ibaraki, Osaka 567}

\vskip 35pt
\hsize= 15.0 true cm
\baselineskip=18pt plus 0.5pt minus 0.1pt
\noindent
\item{}{\bf Abstract: }
The pion-nucleon and kaon-nucleon scattering lengths are studied in the QCD sum
rule.
We show that the leading and next-to-leading order terms of the OPE
give rise to the Tomozawa-Weinberg and sigma terms, respectively.
We also show that in the kaon-nucleon system the $\Lambda(1405)$
contribution has to be subtracted from the OPE side in order to obtain
the scattering length.
The odd components of the $T$-matrices are in agreement with the experimental
values not only in the pion-nucleon channel but also in the kaon-nucleon
channel after the $\Lambda(1405)$ contribution subtracted.
The even components disagree with the experimental values in the pion-nucleon
channel, which is similar to the situation in the PCAC-plus-current-algebra
approach at the Weinberg point.
We speculate that this discrepancy should be explained by the continuum
contribution in the spectral function above the pion-nucleon threshold.

\vfill\eject
\hsize= 16.5 true cm
Recently, it was pointed out that the framework of the QCD sum rule [1,2] can
be
extended to the description of hadronic interactions [3].
In ref. [3] it was shown that the nucleon-nucleon scattering lengths are
related
to the expectation values of the operators such as
$q^{\dagger}q$ and $\bar q q$ with respect to the nucleon.
The calculated scattering lengths are in qualitative agreement
with the experimental values.
The next step we should take is to investigate the interactions of
the nucleon with other hadrons in this formalism.
In the present paper, we apply the formalism to the calculation of the
pion-nucleon and kaon-nucleon scattering lengths.

It is well known that the interactions of the Goldstone bosons with the nucleon
at low energies are determined by the low energy theorems [4].
Namely, the $T$-matrices for the pion-nucleon (kaon-nucleon) scattering can be
calculated by the PCAC and current algebra up to corrections by higher order
terms in the pion (kaon) mass.
For the pion-nucleon system, the low energy theorem is quite successful because
of the small mass of the pion,
and therefore we can check the validity of the QCD sum rules in this
application.
In fact, we will rederive the Tomozawa-Weinberg relation [5,6] as the leading
result
of the QCD sum rules.
The situation is similar to the QCD sum rules for the pion in the vacuum,
where Shifman, Vainshtein and Zakharov [1] rederived the Gell-Mann-Oakes-Renner
relation [7].

For the kaon-nucleon system, however, the low energy theorems are not so
reliable due to the large kaon mass.
Moreover, the experimental results themselves are contradictory: the $K^-$-atom
experiment [8] and the $K^-p$ scattering experiment [9] give the scattering
length which differ from each other in sign.
In the QCD sum rule, we do not assume at least formally that the kaon mass is
small.
Therefore, there is a hope that the QCD sum rule gives better prediction than
the low energy theorem for the kaon-nucleon system and provides some guide to
the resolution of the problem.


We first summarize the derivation of the Borel sum rule for the correlation
function of the axial-vector current for the sake of self-containedness.

Let us consider the correlation function of the axial-vector current,
$$\eqalign{
  \Pi_{\mu\nu}(k) = -i\int {d^4x}e^{ikx}
            \<T(A_\mu(x) A^\dagger_{\nu}(0))\>,
           } \eqno(1)
$$
where the expectation value can be taken with respect to any state and
$$\eqalign{
  A_\mu(x) &= \bar \U(x)\gamma_\mu\gamma_5 \D(x).
            } \eqno(2)
$$
The Lehman representation of the correlation function is given by
$$\eqalign{
  \Pi_{\mu\nu}(\omega,\vec k) = \int {d\omega'}
            {\rho_{\mu\nu}(\omega',\vec k) \over \omega-\omega'}.
           } \eqno(3)
$$
where $\rho^{\mu\nu}(\omega',\vec k)$ is the spectral density:
$$\eqalign{
  \rho_{\mu\nu}(\omega,\vec k)
  =& {i\over2\pi} \left\{\Pi_{\mu\nu}(\omega+i\eta,\vec k)
                        -\Pi_{\mu\nu}(\omega-i\eta,\vec k)\right\}\cr
  =& -{1\over\pi}{\rm Im}\,\Pi_{\mu\nu}(\omega+i\eta,\vec k).
           } \eqno(4)
$$
Hereafter we simply write, e.g.,  ${\rm Im}\,\Pi_{\mu\nu}(\omega,\vec k)$
instead of
${\rm Im}\,\Pi_{\mu\nu}(\omega+i\eta,\vec k)$.
We split $\Pi_{\mu\nu}(\omega,\vec k)$ into even and odd parts:
$$\eqalign{
 \Pi_{\mu\nu}(\omega,\vec k) = \Pi_{\mu\nu{\rm even}}(\omega^2,\vec k)
                 +\omega\Pi_{\mu\nu{\rm odd}}(\omega^2,\vec k).
            } \eqno(5)
$$
Then, eq.(3) can be rewritten in terms of even and odd parts as
$$\eqalign{
 \Pi_{\mu\nu{\rm even}}(\omega^2,\vec k)
  &=-\int_{-\infty}^\infty
{{\rho_{\mu\nu}(\omega',\vec k) \over {\omega'^2-\omega^2}}
 \omega'd\omega'},\cr
 \Pi_{\mu\nu{\rm odd}}(\omega^2,\vec k)
  &=-\int_{-\infty}^\infty
{{\rho_{\mu\nu}(\omega',\vec k) \over {\omega'^2-\omega^2}}
 d\omega'}.
            } \eqno(6)
$$
By applying the Borel transformation,
$$\eqalign{
       L_B\equiv
       \lim_{{n\rightarrow\infty \atop -\omega^2\rightarrow\infty}
       \atop -\omega^2/n = \Borel}
       {(\omega^2)^n\over(n-1)!}\left(-{d\over d\omega^2}\right)^n ,
            } \eqno(7)
$$
to eq.(6) we get
$$\eqalign{
 L_B[\Pi_{\mu\nu{\rm even}}(\omega^2,\vec k)]&=-\int_{-\infty }^\infty
    {\rho_{\mu\nu}(\omega',\vec k)} {\omega'\over \Borel}
    \exp\left(-{\omega'^2 \over \Borel}\right) d\omega',\cr
 L_B[\Pi_{\mu\nu{\rm odd}}(\omega^2,\vec k)]&=-\int_{-\infty }^\infty
    {\rho_{\mu\nu}(\omega',\vec k)} {1\over \Borel}
    \exp\left(-{\omega'^2 \over \Borel}\right) d\omega'
          , } \eqno(8)
$$
where $M_B$ is the Borel mass.
Subtraction terms may be needed in the Lehman representation, eq.(3), but will
disappear after the Borel transformation in eq.(8).
These equations, eq.(8), with the correlation functions on the left-hand side
evaluated by the operator product expansion (OPE) are the Borel sum rules.

Let us next consider the physical content of the spectral function.
The following Ward-Takahashi identity is useful for this purpose:
$$\eqalign{
   &-i \int {d^4x}e^{ikx}k^\mu k^\nu\< T(A_\mu(x) A^\dagger_\nu(0)) \>\cr
 = &-i \int {d^4x}e^{ikx}\big\{
    \<T(\partial^\mu A_\mu(x) \partial^\nu A^\dagger_\nu(0)) \>\cr
   &{\hskip 2cm} +ik_\mu\< \delta(x_0)[A_\mu(x),A^\dagger_0(0)] \>\cr
   &{\hskip 2cm} +\< \delta(x_0)[A_0(x),\partial^\nu A^\dagger_\nu(0)] \>
                           \big\}.\cr
           } \eqno(9)
$$
Since the second and the third terms of the r.h.s are real,
we get the following relation for the imaginary part of eq.(9):
$$\eqalign{
  k^\mu k^\nu{\rm Im}\,\Pi_{\mu\nu}(k)=
  {\rm Im}\left\{-i \int {d^4x}e^{ikx}
    \<T(\partial^\mu A_\mu(x) \partial^\nu A^\dagger_\nu(0)) \>\right\}.
           } \eqno(10)
$$

Hereafter we take $\vec k=0$.
Then only the $\mu=\nu=0$ component of $\Pi_{\mu\nu}$ is relevant.
Therefore, we simplify our notation as follows:
$\Pi(\omega)=\Pi_{00}(\omega,\vec k=0),
 \rho(\omega)=\rho_{00}(\omega,\vec k=0)$.
Thus,
$$\eqalign{
  \rho(\omega) = -{1\over\pi\omega^2}
  {\rm Im}\left\{-i \int {d^4x}e^{i\omega t}
    \<T(\partial^\mu A_\mu(x) \partial^\nu A^\dagger_\nu(0)) \>\right\}.
           } \eqno(11)
$$

We assume that the spectral density in the vacuum, $\rho_0$, is
saturated by the pion (kaon) pole terms:
$$\eqalign{
  \rho_{0}(\omega) = m_\PI f_\PI^2
            \{\delta(\omega-m_\PI)-\delta(\omega+m_\PI)\}.
           } \eqno(12)
$$
It should be noted that $\rho_0$ does not have the pole term due to the
axial-vector meson since $\vec k=0$.

Following ref.[6] we define the off-shell $T$-matrix by
$$\eqalign{
  T(\nu,t,k^2,k'^2)=-i{(k^2-m_\PI^2)(k'^2-m_\PI^2) \over 2f_\PI^2m_\PI^4}
   \int {d^4x}e^{ikx}
   \<N(p)|T(\partial^\mu A_\mu(x) \partial^\nu A^\dagger_\nu(0)) |N(p')\>,
           } \eqno(13)
$$
where $\nu=\omega+t/4M_N$, $t=(k-k')^2$ and $k+p=k'+p'$.
In this and later equations $\<N|{\cal O}|N\>$ means the matrix element
with the disconnected part, $\<0|{\cal O}|0\>\<N|N\>$, subtracted.
$f_\PI$ is defined in the standard way:
$$\eqalign{
  \<0|A_\mu(0)|\PI(k)\> = i\sqrt{2}f_\PI k_\mu.
           } \eqno(14)
$$
In this definition the $T$-matrix is related to the $S$-matrix by
$S_{fi}=\delta_{fi}-i(2\pi)^4\delta^4(k'+p'-k-p)T_{fi}$.

{}From eqs.(11) and (13) the spectral density, in which the expectation value
is taken
with respect to the nucleon, becomes
$$\eqalign{
  \rho_{N}(\omega)
  =&{-1 \over \pi\omega^2}{\rm Im}
   \left\{{2f_\PI^2m_\PI^4 \over (\omega^2-m_\PI^2)^2}
   T(\omega,0,\omega^2,\omega^2)\right\}\cr
  =&-{1 \over 2}f_\PI^2
   \bigg[\delta'(\omega-m_\PI){\rm Re}T_+
   -\delta(\omega-m_\PI){\rm Re}\left(T'_+
   -{3 \over m_\PI}T_+\right)\cr
  &+\delta'(\omega+m_\PI){\rm Re}T_-
   +\delta(\omega+m_\PI){\rm Re}\left(T'_-
   -{3 \over m_\PI}T_-\right)\cr
  &+{4m_\PI^4 \over \omega^2}
   {\rm Re}{1 \over (\omega^2-m_\PI^2)^2}
   {1 \over \pi}{\rm Im}\,T(\omega,0,\omega^2,\omega^2)
     \bigg],\cr
           } \eqno(15)
$$
where $T_\pm=T(\pm m_\PI,0,m_\PI^2,m_\PI^2)$,
$T'_\pm=\pm {\partial \over \partial \omega}
T(\omega,0,\omega^2,\omega^2)|_{\omega=\pm m_\PI}$.
The last term in eq.(15) represents the contribution from the continuum
scattering states above the pion-nucleon (kaon-nucleon) threshold as well
as bound states below the threshold if they exist.
In this paper we ignore the former.
The later  will be considered later when it becomes necessary.
For the time being we do not explicitly write down the bound state
contributions for notational simplicity.
Thus, we take the spectral function to be
$$\eqalign{
  \rho_{N}(\omega)
  =&-{1 \over 2}f_\PI^2
   \bigg[\delta'(\omega-m_\PI){\rm Re}T_+
   -\delta(\omega-m_\PI){\rm Re}\left(T'_+
   -{3 \over m_\PI}T_+\right)\cr
  &\qquad\qquad
   +\delta'(\omega+m_\PI){\rm Re}T_-
   +\delta(\omega+m_\PI){\rm Re}\left(T'_-
   -{3 \over m_\PI}T_-\right)\bigg].\cr
           } \eqno(16)
$$

By substituting eqs.(12) and (16) into eq.(8) we get
$$\eqalign{
  -2{m_\PI^2 \over M_B^2} f_\PI^2 \exp\left(-{m_\PI^2 \over \Borel}\right) =
  L_B[\Pi_{0 \rm even}(\omega^2)],
           } \eqno(17)
$$
$$\eqalign{
 -2{m_\PI^2 \over M_B^6} f_\PI^2 T_{\pm}
 \exp\left(-{m_\PI^2 \over \Borel}\right)
 =L_B'[\Pi_{N \rm even}(\omega^2)] \pm m_\PI L_B'[\Pi_{N \rm odd}(\omega^2)],
           } \eqno(18)
$$
$$\eqalign{
 -2{m_\PI^3 \over M_B^6} f_\PI^2 T'_{\pm}
 \exp\left(-{m_\PI^2 \over \Borel}\right)
 =2L_B''[\Pi_{N \rm even}(\omega^2)] \pm 3m_\PI L_B'''[\Pi_{N \rm odd}
(\omega^2)],
           } \eqno(19)
$$
where
$$\eqalign{
  L_B'[\Pi_N(\omega^2)] =&
            {d \over d\Borel}L_B[\Pi_N(\omega^2)] \cr
    &+\left(1-{m_\PI^2 \over \Borel}\right){1 \over \Borel}
            L_B[\Pi_N(\omega^2)],\cr
  L_B''[\Pi_N(\omega^2)] =&
     \left(1+{m_\PI^2\over\Borel}\right)
       {d \over d\Borel}L_B[\Pi_N(\omega^2)] \cr
    &+\left(1+{m_\PI^2 \over \Borel}
              -{m_\PI^4 \over \Borels}\right){1 \over \Borel}
       L_B[\Pi_N(\omega^2)], \cr
  L_B'''[\Pi_N(\omega^2)] =&
     \left(1+{2 \over 3}{m_\PI^2\over\Borel}\right)
       {d \over d\Borel}L_B[\Pi_N(\omega^2)] \cr
    &+\left(1+{1 \over 3}{m_\PI^2 \over \Borel}
             -{2 \over 3}{m_\PI^4 \over \Borels}\right){1 \over \Borel}
       L_B[\Pi_N(\omega^2)].
           } \eqno(20)
$$

{}From eqs.(17), (18) and (19) $T_\pm$ and $T'_\pm$ are given by
$$\eqalign{
  T_{\pm} = \Borels
  {L_B'[\Pi_{N\,{\rm even}}(\omega^2)]
   \pm m_\PI L_B'[\Pi_{N\,{\rm odd}}(\omega^2)]
   \over  L_B[\Pi_{0\,{\rm even}}(\omega^2)]},
           } \eqno(21)
$$
$$\eqalign{
  T'_{\pm} = {\Borels \over m_\PI}
  {2L_B''[\Pi_{N\,{\rm even}}(\omega^2)]
   \pm 3m_\PI L_B'''[\Pi_{N\,{\rm odd}}(\omega^2)]
   \over  L_B[\Pi_{0\,{\rm even}}(\omega^2)]}.
           } \eqno(22)
$$

Let us now turn to the OPE.
The leading and next-to-leading order terms of the correlation function in the
OPE can be read off from the Ward-Takahashi identity, eq.(9), without
explicitly performing the OPE.
We rewrite eq.(9) as
$$\eqalign{
   &-i \int {d^4x}e^{ikx}k^\mu k^\nu\< T(A_\mu(x) A^\dagger_\nu(0)) \>\cr
= & k^\mu\<\bar q_1\gamma_\mu q_1 - \bar q_2\gamma_\mu q_2\>
  -(m_1+m_2)\<\bar q_1 q_1 + \bar q_2 q_2\>\cr
 &-(m_1+m_2)^2 i\int d^4xe^{ikx}
 \<T\left(\varphi(x) \varphi^\dagger(0)\right)\>,
           } \eqno(23)
$$
where
$$\eqalign{
 \varphi(x)=i\bar q_1(x)\gamma_5 q_2(x).
           } \eqno(24)
$$
On the right-hand side of eq.(23), the dimensions of the operators in the
first,
second and third terms are three, four and five or higher, respectively.
(The OPE for the correlation function of $\varphi$ has at least dimension
three.)
Moreover, their quark mass dependence is constant, linear and quadratic,
respectively.
Therefore, the first term is the most important, the second is next and the
third is the least important not only in the sense of the OPE but also in the
sense of the chiral symmetry breaking expansion.
Thus, we first concentrate on the first two terms in eq.(23).
The effect of the higher dimension operators in the last term will be
discussed later.

In ref.~[1], Shifman, Vainshtein and Zakharov showed that
from the sum rules in the vacuum, essentially eqs.(17) and (23),
the Gell-Mann-Oakes-Renner relation [7] is rederived
$$\eqalign{
  2 m_\PI^2 f_\PI^2
  = -\left(m_\u+m_\d\right)\<\bar \U \U+\bar \D \D\>_0,
           } \eqno(25)
$$
where higher order terms of $m_\pi^2/M_B^2$ are neglected.

Similarly, we can show that from the sum rules in the nucleon, eqs.(21), (22)
and
(23), the following relations for the $T$-matrix and its derivative are
derived:
$$\eqalign{
  T_{\PI N} = -{m_\PI\<q_1^\dagger q_1-q_2^\dagger q_2\>_N
               -(m_1+m_2)\<\bar q_1q_1+\bar q_2q_2\>_N \over
                2f_\PI^2},
           } \eqno(26)
$$
$$\eqalign{
  T'_{\PI N} = -{\<q_1^\dagger q_1-q_2^\dagger q_2\>_N \over
                2f_\PI^2}.
           } \eqno(27)
$$

The matrix elements of the operators in eqs.(26) and (27),
$\<q^\dagger q\>_N$ and $m_q\<\bar qq\>_N$, are given by the quark
number in the nucleon and the nucleon sigma term, respectively:
$$\eqalign{
&\<u^\dagger u\>_p=\<d^\dagger d\>_n=2,\cr
&\<d^\dagger d\>_p=\<u^\dagger u\>_n=1,\cr
&\<s^\dagger s\>_p=\<s^\dagger s\>_n=0,\cr
           } \eqno(28)
$$
and
$$\eqalign{
	&{m_u+m_d\over 2}\<\bar uu+\bar dd\>_p
	={m_u+m_d\over 2}\<\bar uu+\bar dd\>_n
	=\sigma_{\pi N},\cr
	&{m_u+m_s\over 2}\<\bar uu+\bar ss\>_p
        ={m_s+m_d\over 2}\<\bar ss+\bar dd\>_n
	=\sigma_{K p},\cr
	&{m_u+m_s\over 2}\<\bar uu+\bar ss\>_n
	={m_s+m_d\over 2}\<\bar ss+\bar dd\>_p
	=\sigma_{K n}.
           } \eqno(29)
$$
We use the following values for the sigma terms in later calculations:
$\sigma_{\pi N}=45 {\rm MeV}$, $\sigma_{K p}=374 {\rm MeV}$ and
$\sigma_{K n}=330 {\rm MeV}$.
The pion-nucleon sigma term is taken from ref.[10] and the kaon-nucleon
sigma terms are calculated from the above pion-nucleon sigma term,
the quark masses, $m_u=m_d=7\;{\rm MeV}$, $m_s=170\;{\rm MeV}$
and the $y$ parameter given in ref.[10],
$y=2\<\bar ss\>_p/\<\bar uu+\bar dd\>_p=0.2$.

Let us first look at the pion-nucleon channel.
The $T$-matrices and their derivatives are given by
$$\eqalign{
  T^{(+)}_{\pi N}&={(m_u+m_d)\<\bar uu+\bar dd\>_N
                   \over 2 f_\pi^2}
                  ={\sigma_{\pi N}\over f_\pi^2}, \cr
  T^{(-)}_{\pi N}&=-{m_\pi\<u^\dagger u-d^\dagger d\>_p
                   \over 2 f_\pi^2}
                  =-{m_\pi\over 2f_\pi^2},\cr
  T'^{(+)}_{\pi N}&=0, \cr
  T'^{(-)}_{\pi N}&=-{\<u^\dagger u-d^\dagger d\>_p \over 2f_\pi^2}
                   =-{1 \over 2f_\pi^2},
           } \eqno(30)
$$
where $T_{\pi N}^{(\pm)}={1\over2}(T_{\pi^- p}\pm T_{\pi^+ p})
={1\over2}(T_{\pi^+ n}\pm T_{\pi^- n})$.

The leading order term, the dimension-three operator, contributes to the
isospin-odd component of the $T$-matrix and gives the Tomozawa-Weinberg term
[5,6].
The next-to-leading order term, the dimension-four operator, contributes to
the isospin-even component of the $T$-matrix and gives the sigma term, which is
the same as that obtained by using the PCAC and current algebra at the
Weinberg point [11].

It is interesting that the quark number in the nucleon determines the
leading-order
form of the $T$-matrix.
It should be emphasized again that in the present approach the $T$-matrix is
obtained at the pion-nucleon threshold,
$\nu=m_\pi$, $t=0$, $k^2=k'^2=m_\pi^2$, while in the PCAC-plus-current-algebra
 approach it is at the Weinberg point, i.e. $\nu=t=k^2=k'^2=0$.

The calculated scattering lengths are tabulated in table 1, where the observed
scattering lengths are also shown for comparison.
The scattering lengths calculated with the dimension-three operator, the
entries
in the first column, are surprisingly close to the experimental
values, which is exactly the same result as in refs.[5,6].
The contribution of the dimension-four operator in the isospin-even component
of the $T$-matrix, the entry in the second column, is much larger than the
experimental value.
This is similar to the fact that the $T$-matrix at the Weinberg point in the
PCAC-plus-current-algebra approach is much larger than the experimental value
at the pion-nucleon threshold.
The latter difference is usually attributed to sources such as resonance
and/or smooth background contributions, which cannot be determined by the
symmetry argument alone.
Therefore, a natural question is what cancels the sigma term contribution
to the $T$-matrix at the pion-nucleon threshold in the present approach.
There are two possibilities: one is higher order terms in the OPE and the
other is the continuum contribution above the pion-nucleon threshold in the
spectral function.
We will come back to this point later.
\midinsert
\noindent
\item{}{\bf Table 1:}
Calculated and observed pion-nucleon and kaon-nucleon scattering lengths in the
unit of fm.
Experimental values are taken from ref.[12] for the pion-nucleon channel and
ref.[13] for the kaon-nucleon channel.
In the parentheses are shown the scattering lengths calculated without
$\Lambda(1405)$ contribution.
We explicitly showed the errors in the calculated scattering lengths due to the
$\bar K N \Lambda(1405)$ coupling constant.
Certainly, there are other errors in the calculated and observed scattering
lengths.
However, we cannot specify explicit numbers for those errors except for
the observed pion-nucleon scattering lengths, which are small anyway.
\vskip 5mm
{\offinterlineskip
 \halign{\strut\vrule#&\quad\hfil\bf#\hfil\quad&&
        \vrule#&\quad#\hfil\quad\cr
         \noalign{\hrule}
         \noalign{\hrule}
  &{  }
    && \tallbox{2}\hfil{dim. 3}
    && \hfil{$\leq$ dim. 4}
    && \hfil{$\leq$ dim. 6}
    && \hfil{Experiment}
  &\cr
         \noalign{\hrule}
  &\tallbox{2}{$a_{\pi N}^{(+)}$}
    && \hfil{$0$}  && \hfil{$-0.07$}
    && \hfil{$-0.07$}  && \hfil{$-0.01$}
  &\cr
  &\tallbox{2}{$a_{\pi N}^{(-)}$}
    && \hfil{$0.11$}   && \hfil{$0.11$}
    && \hfil{$0.12$}   && \hfil{$0.13$}
  &\cr
  &\tallbox{2}{$a_{K p}^{(+)}$}
    && \hfil{$-0.67\pm0.54$}   && \hfil{$-0.97\pm0.54$}
    && \hfil{$-1.02\pm0.54$}   && \hfil{$-0.50$}
  &\cr
  &\tallbox{3}{ }
    && \hfil{($0$)}          && \hfil{($-0.29$)}
    && \hfil{($-0.34$)} && \hfil{ }
  &\cr
  &\tallbox{2}{$a_{K p}^{(-)}$}
    && \hfil{$-0.28\pm0.54$}   && \hfil{$-0.28\pm0.54$}
    && \hfil{$-0.22\pm0.54$}   && \hfil{$-0.17$}
  &\cr
  &\tallbox{3}{ }
    && \hfil{($0.39$)}         && \hfil{($0.39$)}
    && \hfil{($0.45$)}         && \hfil{ }
  &\cr
  &\tallbox{2}{$a_{K n}^{(+)}$}
    && \hfil{$0$}              && \hfil{$-0.26$}
    && \hfil{$-0.31$}          && \hfil{$0.10$}
  &\cr
  &\tallbox{2}{$a_{K n}^{(-)}$}
    && \hfil{$0.20$}   && \hfil{$0.20$}
    && \hfil{$0.23$}   && \hfil{$0.27$}
  &\cr
         \noalign{\hrule}
 } }
\endinsert

Let us turn to the kaon-nucleon channel.
Since $\Lambda(1405)$ exists below the $\bar K N$ threshold, we should take
into account its contribution to the spectral function.
This is done by the following replacement [9]:
$$\eqalign{
  &T_{K^-p}\rightarrow
         T_{K^-p}-{g_{\Lambda^*}^2}{m_K^2\over m_K+M_N-M_{\Lambda^*}}
         \left({1\over M_{\Lambda^*}-M_N}\right)^2, \cr
  &T_{K^-p}'\rightarrow T_{K^-p}'
                 +{g_{\Lambda^*}^2}{m_K\over (M_{\Lambda^*}-M_N)^2}
                   \left({1\over M_{\Lambda^*}-M_N-m_K}\right)^2
                   \left\{ 2(M_{\Lambda^*}-M_N)-m_K \right\},
           } \eqno(31)
$$
where $g_{\Lambda^*}$ is the $\bar K N \Lambda(1405)$ coupling constant,
$M_{\Lambda^*}$ is the mass of $\Lambda(1405)$.
In eq.~(31) we neglected higher order terms in the binding energy,
$m_K+M_N-M_{\Lambda^*}$.
Having done this modification, we obtain the $T$-matrices and their derivatives
as
$$\eqalign{
  T^{(+)}_{Kp}&={\sigma_{Kp}\over f_K^2}
               +{g_{\Lambda^*}^2\over2}{m_K^2\over m_K+M_N-M_{\Lambda^*}}
                \left({1\over M_{\Lambda^*}-M_N}\right)^2,\cr
  T^{(+)}_{Kn}&={\sigma_{Kn}\over f_K^2}, \cr
  T^{(-)}_{Kp}&=-{m_K\over f_K^2}
               +{g_{\Lambda^*}^2\over2}{m_K^2\over m_K+M_N-M_{\Lambda^*}}
                \left({1\over M_{\Lambda^*}-M_N}\right)^2,\cr
  T^{(-)}_{Kn}&=-{m_K\over 2f_K^2},\cr
  T'^{(+)}_{Kp}&=-{g_{\Lambda^*}^2\over2}{m_K\over (M_{\Lambda^*}-M_N)^2}
                  \left({1\over M_{\Lambda^*}-M_N-m_K}\right)^2
                  \left\{ 2(M_{\Lambda^*}-M_N)-m_K \right\},\cr
  T'^{(+)}_{Kn}&=0,\cr
  T'^{(-)}_{Kp}&=-{1\over f_K^2}
                 -{g_{\Lambda^*}^2\over2}{m_K\over (M_{\Lambda^*}-M_N)^2}
                   \left({1\over M_{\Lambda^*}-M_N-m_K}\right)^2
                   \left\{ 2(M_{\Lambda^*}-M_N)-m_K) \right\},\cr
  T'^{(-)}_{Kn}&=-{1\over 2f_K^2},
           } \eqno(32)
$$
where $T_{K N}^{(\pm)}={1\over2}(T_{K^- N}\pm T_{K^+ N})$.

The results for the kaon-nucleon channel are similar to those for the
pion-nucleon
channel except for the contribution from $\Lambda(1405)$:
the dimension-three operator gives the Tomozawa-Weinberg term to $T^{(-)}$ and
the dimension-four operator gives the sigma term to $T^{(+)}$.

The calculated and observed scattering lengths are tabulated in table 1.
If we believe the experimental values for the kaon-nucleon scattering lengths,
the original Tomozawa-Weinberg results, the entries in the first column for
$a_{Kn}^{(\pm)}$ and those in the parentheses in the first column for
$a_{Kn}^{(\pm)}$, are not so successful as in the pion-nucleon channel.
In the present approach, the leading terms are those with the contribution
from $\Lambda(1405)$ included.
Though the error of the $\Lambda(1405)$ contribution due to the
experimental uncertainty of the $\bar{K}N\Lambda(1405)$ coupling constant
is quite large, the
inclusion of the $\Lambda(1405)$ contribution seems to improve the agreement
with the observed values.
However, we should also keep in mind that the experimental situation concerning
the $K^-$$p$ scattering length is still controversial.
Namely, the scattering lengths determined by the atomic experiment, ref.~[8],
differ in sign from those shown in table 1, which are determined by the
scattering experiment.
Sound experimental determination of the scattering length as well as the
$\bar{K}N\Lambda(1405)$ coupling constant is needed.

Let us now consider the effect of the higher order terms of the OPE.
The OPE of the correlation function is given up to dimension six as
follows:
$$\eqalign{
 \Pi_{0\rm odd}(\omega^2) =& 0,\cr
 \Pi_{0\rm even}(\omega^2) =&
   {3\over8\pi^2}(m_\u+m_\d)^2\ln(-\omega^2)
  -\left(m_\u+m_\d\right)\<\bar \U\U+\bar \D\D\>_0
    {1 \over \omega^2}\cr
 &+{1\over8} (m_\u+m_\d)^2
    \<{\alpha_s\over\pi}G_{\mu\nu}G^{\mu\nu}\>_N{1\over\omega^4},\cr
 \Pi_{N\rm odd}(\omega^2) =&
   \<\U^\dagger \U-\D^\dagger \D\>_N{1\over\omega^2}
  +(m_\u+m_\d)^2
   \<\U^\dagger \U-\D^\dagger \D\>_N{1\over\omega^4},\cr
 \Pi_{N\rm even}(\omega^2) =&
  -(m_\u+m_\d)\<\bar \U\U+\bar \D\D\>_N{1\over\omega^2}\cr
 &-(m_\u+m_\d)^2\Big\{2\left(i\<\bar \U{\cal S}(\gamma_0D_0)\U\>_N
                            +i\<\bar \D{\cal S}(\gamma_0D_0)\D\>_N\right)\cr
 &\qquad -{1\over8}\<{\alpha_s\over\pi}G_{\mu\nu}G^{\mu\nu}\>_N
         -{1\over2}\<{\alpha_s\over\pi}
           {\cal S}(G_{0\nu}G_0^{\nu})\>_N\Big\}{1\over\omega^4},
          } \eqno(33)
$$
where ${\cal S}[A_\mu B_\nu]$ means the symmetric and traceless tensor
made of $A_\mu$ and $B_\nu$.
By substituting eq.(33) into eq.(17) we obtain
$$\eqalign{
  2 m_\PI^2 f_\PI^2 {1\over\Borel}\exp\left(-{m_\PI^2 \over \Borel}\right)
  =& -(m_\u+m_\d)\<\bar \U\U+\bar \D\D\>_0{1\over\Borel}\cr
   & +(m_\u+m_\d)^2\left\{{3\over8\pi^2}-{1\over8}
      \<{\alpha_s\over\pi}G_{\mu\nu}G^{\mu\nu}\>_0{1\over\Borels}\right\}.
           } \eqno(34)
$$
Similarly, by substituting eq.(33) into eqs.(21) and (22) and dividing them by
eq.(34) we obtain
$$\eqalign{
  &T_{\PI N}= \cr
  &{m_\PI^2\over\Borel}{  m_\PI\< \U^\dagger \U - \D^\dagger \D \>_N
      -(m_\u+m_\d)\< \bar\U\U + \bar\D\D\ \>_N
      -\< \tilde{\cal O}_1 \>_N
       \left({1\over m_\PI^2}-{1\over\Borel}\right)
      \over
    -{3\over8\pi^2}(m_\u+m_\d)^2
    +\left(m_\u+m_\d\right)\< \bar\U\U + \bar\D\D\ \>_0
          {1\over\Borel}
    +(m_\u+m_\d)^2{1\over8}\<{\alpha_s\over\pi} G_{\mu\nu}G^{\mu\nu}\>_0
          {1\over\Borels} },
           } \eqno(35)
$$
$$\eqalign{
  &T'_{\PI N}=\cr
  &{m_\PI\over\Borel} { m_\PI\< \U^\dagger \U - \D^\dagger \D \>_N
     \left(1+2{m_\PI^2\over\Borel}\right)
      -2(m_\u+m_\d)(\bar\U\U + \bar\D\D){m_\PI^2\over\Borel}
      -\<\tilde{\cal O}_2 \>_N
       \left({1\over m_\PI^2}+{1\over\Borel}\right)
      \over
     -{3\over8\pi^2}(m_\u+m_\d)^2
     +\left(m_\u+m_\d\right)\< \bar\U\U + \bar\D\D\ \>_0
          {1\over\Borel}
     +(m_\u+m_\d)^2{1\over8}\<{\alpha_s\over\pi} G_{\mu\nu}G^{\mu\nu}\>_0
          {1\over\Borels} },
           } \eqno(36)
$$
where
$$\eqalign{
  \tilde{\cal O}_1=&(m_\u+m_\d)^2\Big[
   m_\PI \left(\U^\dagger \U - \D^\dagger \D \right)\cr
    &- \Big\{2\left(i\bar\U{\cal S}(\gamma_0D_0)\U
                           + i\bar\D{\cal S}(\gamma_0D_0)\D\right)
     -{1\over8}{\alpha_s\over\pi}G_{\mu\nu}G^{\mu\nu}
     -{1\over2}{\alpha_s\over\pi}{\cal S}(G_{0\nu}G_0^{\nu})\Big\}\Big] ,
          } \eqno(37)
$$
$$\eqalign{
 \tilde{\cal O}_2=&(m_\u+m_\d)^2\Big[
   3m_\PI \left(\U^\dagger \U - \D^\dagger \D \right)\cr
    &- 2\Big\{2\left(i\bar\U{\cal S}(\gamma_0D_0)\U
                           + i\bar\D{\cal S}(\gamma_0D_0)\D\right)
     -{1\over8}{\alpha_s\over\pi}G_{\mu\nu}G^{\mu\nu}
     -{1\over2}{\alpha_s\over\pi}{\cal S}(G_{0\nu}G_0^{\nu})\Big\} \Big].
          } \eqno(38)
$$

The scattering lengths with higher order terms included are shown in the third
column
of table 1.
The matrix elements of the operators with respect to the nucleon are taken
from ref.[14], and are
$$\eqalign{
  &i\<{\cal S}[\bar u\gamma_\mu D_\nu u]\>_p =
   i\<{\cal S}[\bar d\gamma_\mu D_\nu d]\>_n = 222 \;{\rm MeV},\cr
  &i\<{\cal S}[\bar d\gamma_\mu D_\nu d]\>_p =
   i\<{\cal S}[\bar u\gamma_\mu D_\nu u]\>_n = 95 \;{\rm MeV},\cr
  &i\<{\cal S}[\bar s\gamma_\mu D_\nu s]\>_p =
   i\<{\cal S}[\bar s\gamma_\mu D_\nu s]\>_n = 18 \;{\rm MeV},\cr
  &\<{\alpha_s\over\pi}G_{\mu\nu}G^{\mu\nu}\>_N =-738 \;{\rm MeV},\qquad
   \<{\alpha_s\over\pi}{\cal S}[G_{\mu 0}G^{\mu 0}]\>_N =-50 \;{\rm MeV}.
           }
$$
The condensates of the operators in the vacuum are taken from ref.[2]
and are
$$\eqalign{
&\<\bar uu\>_0 =\<\bar dd\>_0 = -(225\;{\rm MeV})^3,\cr
&\<\bar ss\>_0 = -(217\;{\rm MeV})^3, \cr
&\<{\alpha_s\over\pi}G^2\>_0 = (340\;{\rm MeV})^4.
           }
$$
It should be noted that these matrix elements both with the nucleon and
the vacuum are for the renormalization scale, $1\;{\rm GeV}$.
The Borel mass is taken to be $M_B^2=1\;{\rm GeV}^2$.
The calculated scattering lengths, however, are almost constant in the range,
$\Borel=1\sim1.5\;{\rm GeV}^2$.

In the pion-nucleon channel the effect of the dimension-five and six operators
is very small.
Also in the kaon-nucleon channel the effect is small but not so extreme as in
the pion-nucleon channel.
This is because the kaon mass is not so small as the pion mass.

We now come back to the question concerning the discrepancy in the
isospin-even component of the $T$-matrix in the pion-nucleon channel.
Since we have observed that the effect of the dimension-five and six terms of
the OPE is small, it is unlikely that the difference is explained by higher
order terms of the OPE.
Therefore, we speculate that the continuum contribution near above the
pion-nucleon threshold should be responsible for the discrepancy.
This is our future problem.

Finally, we would like to mention the physical significance of $T'$, the
derivative of the $T$-matrix with respect to the pion energy, $\omega$.
In the sum rule approach, only on-shell quantities appear in the spectral
function.
Therefore, off-shell $T$-matrices, which are important when one discusses the
pion (kaon) mass in nuclear matter [15], cannot be directly obtained.
We can, however, obtain $T'$, the slope of the $T$-matrix as a function of
the pion energy, at on-shell points, with which we can linearly extrapolate the
$T$-matrices from the on-shell to off-shell points.
Moreover, $T'$ turns out to be related to the change of the decay constant
of the pion (kaon) in nuclear matter.
These will be discussed in detail in our next paper [16].

In summary, we studied the pion-nucleon and kaon-nucleon scattering
lengths in the QCD sum rule.
We showed that the leading and next-to-leading order terms of the OPE
give rise to the Tomozawa-Weinberg and sigma terms, respectively.
The effect of the higher order terms is extremely small in the pion-nucleon
channel and small in the kaon-nucleon channel.
It was also shown that in the kaon-nucleon channel the $\Lambda(1405)$
contribution has to be subtracted from the OPE side in order to obtain
the scattering length.
The odd components of the $T$-matrices are in agreement with the experimental
values not only in the pion-nucleon channel but also in the kaon-nucleon
channel after the $\Lambda(1405)$ contribution subtracted.
The even components in the pion-nucleon channel disagree with the experimental
values due to the large sigma term, which is similar to the situation in the
PCAC-plus-current-algebra approach at the Weinberg point.
We expect that this discrepancy should be explained by the continuum
contribution above the pion-nucleon threshold.

We would like to thank Professor H.~Terazawa for discussions concerning low
energy theorems and Dr. T.~Muto for bringing our attention to the derivatives
of the pion-nucleon and kaon-nucleon $T$-matrices.
We would also like to acknowledge Dr. M.~Arima for explaining to us the
determination of the $\bar K N \Lambda(1405)$ coupling constant.

\vfill\eject
\vskip 20pt
\leftline{\bf References}
\vskip 10pt
\item{[1]} M.A.~Shifman, A.I.~Vainshtein and V.I. Zakharov,
                Nucl. Phys. {\bf B147} (1979) 385.
\item{[2]} L.J.~Reinders, H.~Rubinstein and S.~Yazaki,
                Phys. Rep. {\bf 127} (1985) 1, and references therein.
\item{[3]} Y.~Kondo and  O.~Morimatsu,
                Phys. Rev. Lett. {\bf 71} (1993) 2855.
\item{[4]} For a review see S.L. Adler and R.F. Dashen,
                Current algebra and applications (Benjamin, New York, 1968).
\item{[5]} Y.~Tomozawa, Nuovo Cimento {\bf 46A} (1966) 707.
\item{[6]} S. Weinberg, Phys. Rev. Lett. {\bf 17} (1966) 616.
\item{[7]} M.~Gell-Mann, R.~J.~Oakes and B.~Renner, Phys. Rev. {\bf 175} (1968)
2195.
\item{[8]} J.~D.~Davis et al., Phys. Lett. {\bf B83} (1979) 55,
\item{   } M.~Izycki et al., Z. Phys. {\bf A297} (1980) 11,
\item{   } P.~M.~Bird et al., Nucl. Phys. {\bf A404} (1983) 482.
\item{[9]} O.~Dumbrajs et al., Nucl. Phys. {\bf B216} (1983) 277.
\item{[10]} J.~Gasser, H.~Leutwyler and M.~E.~Sainio, Phys. Lett. {\bf B253}
(1991) 252.
\item{[11]} T.~P.~Cheng and R.~Dashen, Phys. Rev. Lett. {\bf 26} (1971) 594.
\item{[12]} R.~Koch and E.~Pietarinen, Nucl. Phys. {\bf A336} (1980) 331.
\item{[13]} A.~D.~Martin, Nucl. Phys. {\bf B179} (1981) 33.
\item{[14]} E.D.~Drukarev and E.M.~Levin,
        Nucl. Phys. {\bf A511} (1990) 679, {\bf A516} (1990) 715(E),
\item{   } T.~Hatsuda and S.H.~Lee, Phys. Rev. {\bf C 46} (1992) R34,
\item{   } X.~Jin, T.D.~Cohen, R.J.~Furnstahl and D.K.~Griegel,
               Phys. Rev. {\bf C 47} (1993) 2882.
\item{[15]} A.~E.~Nelson and D.~B.~Kaplan, Phys. Lett. {\bf B192} (1984) 193,
\item{   }  G.~E.~Brown, V.~Koch and M.~Rho, Nucl. Phys. {\bf A535} (1991) 701,
\item{   }  J.~Delorme, M.~Ericson and T.~E.~O.~Ericson, Phys. Lett.
            {\bf B291} (1992) 379,
\item{   }  H.~Yabu, S.~Nakamura and K.~Kubodera, Phys. Lett. {\bf B317} (1993)
            269,
\item{   }  H.~Yabu, S.~Nakamura, F.~Myhrer and K.~Kubodera,
            Phys. Lett. {\bf B315} (1993) 17. 

\item{[16]} Y.~Kondo, O.~Morimatsu and Y.~Nishino, in preparation.

\end